\documentclass[aps,prl,twocolumn,superscriptaddress,floatfix,showpacs,longbibliography,10pt]{revtex4-2}
\usepackage[T1]{fontenc}
\usepackage{graphicx,color}
\usepackage[utf8]{inputenc}
\usepackage{amsmath,amssymb,bm,amsfonts,dsfont,mathrsfs,amsthm}
\usepackage{braket}
\usepackage{txfonts,comment}
\usepackage[colorlinks=true,linkcolor=blue,citecolor=blue,urlcolor=blue]{hyperref}
\usepackage{soul}
\usepackage{tikz}
\usepackage{physics}
\usetikzlibrary{shapes, arrows, positioning, calc, backgrounds}

\usepackage{dcolumn}
\usepackage{bm}

\usepackage{hyperref}
\usepackage{xr-hyper}
\usepackage{algorithm}
\usepackage{algpseudocode}
\usepackage{xcolor}

\begin{document}

\preprint{APS/123-QED}

\title{Tensor Network Assisted Distributed Variational Quantum Algorithm for Large Scale Combinatorial Optimization Problem}








\author{Yuhan Huang}
\email{yhuangfv@ust.hk}
\affiliation{The Department of Electronic and Computer Engineering, The Hong Kong University of Science and Technology, 999077, Hong Kong}

\author{Siyuan Jin}
\affiliation{The Department of Information Systems, The Hong Kong University of Science and Technology, 999077, Hong Kong}

\author{Yichi Zhang}
\affiliation{Department of Physics, The Hong Kong University of Science and Technology,
999077, Hong Kong}

\author{Qi Zhao}
\affiliation{The Department of Computer Science, The University of Hong Kong, Hong Kong}

\author{Jun Qi}
\affiliation{The Department of Computer Science, Hong Kong Baptist University, Hong Kong}

\author{Qiming Shao}
\email{eeqshao@ust.hk}
\affiliation{The Department of Electronic and Computer Engineering, The Hong Kong University of Science and Technology, 999077, Hong Kong}
\affiliation{Department of Physics, The Hong Kong University of Science and Technology,
999077, Hong Kong}

\begin{abstract}
\noindent
Although quantum computing holds promise for solving Combinatorial Optimization Problems (COPs), the limited qubit capacity of NISQ hardware makes large-scale instances intractable. Conventional methods attempt to bridge this gap through decomposition or compression, yet they frequently fail to capture global correlations of subsystems, leading to solutions of limited quality. We propose the Distributed Variational Quantum Algorithm (DVQA) to overcome these limitations, enabling the solution of 1,000-variable instances on constrained hardware. A key innovation of DVQA is its use of the truncated higher-order singular value decomposition to preserve inter-variable dependencies without relying on complex long-range entanglement, leading to a natural form of noise localization where errors scale with subsystem size rather than total qubit count, thus reconciling scalability with accuracy. Theoretical bounds confirm the algorithm's robustness for $p$-local Hamiltonians. Empirically, DVQA achieves state-of-the-art performance in simulations and has been experimentally validated on the “Wu Kong” quantum computer for portfolio optimization. This work provides a scalable, noise-resilient framework that advances the timeline for practical quantum optimization algorithms.

\end{abstract}

 \maketitle

\begin{figure*}[htpb]
    \centering
    \includegraphics[width=\linewidth]{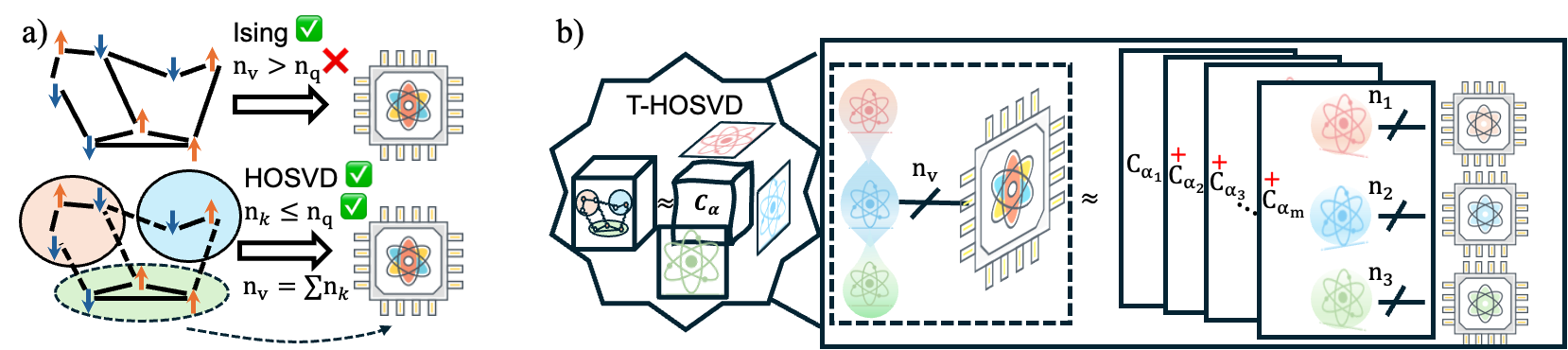}
    \caption{a) Large-scale COPs with $n_v$ decision variables often require more qubits than are available on today’s quantum processing units, i.e., $n_v > n_q$, where $n_q$ denotes the number of physical qubits. To address this limitation, the problem can be partitioned into smaller subsystems with $n_k$ qubits and transformed using higher-order singular value decomposition (HOSVD), enabling the solution of tractable subproblems on limited hardware. b)  Specifically, truncated HOSVD (T-HOSVD) factorizes the global coefficient tensor into lower-rank components, each of which can be executed on a smaller QPU. The global solution is then reconstructed through variational recombination of the subproblem results.}
    \label{fig1:motivation}
\end{figure*}

\emph{\textbf{Introduction.--}} 
Combinatorial optimization problems (COPs)~\cite{jordan2025optimization} span diverse domains, including finance~\cite{mugel2022dynamic,mcmahon2024improving,huber2024exponential,deshpande2025currency}, graph theory~\cite{lykov2023sampling,farhi2014quantum}, and physics~\cite{pelofske2024short,heim2015quantum,layden2023quantum}. While algorithms like quantum annealing (QA)~\cite{morita2008mathematical,morrell2024quantumannealing}, quantum-enhanced markov chain monte carlo~\cite{layden2023quantum}, quantum approximate optimization algorithm (QAOA)~\cite{farhi2014quantum,blekos2024review}, and variational quantum eigensolver (VQE)~\cite{kandala2017hardware,cao2024accelerated,hou2025effcient,cerezo2021variational} offer pathways to solve these problems in the NISQ era~\cite{preskill2018quantum}, their practical implementation faces significant hurdles. Specifically, the scalability of these algorithms is stifled by the dual bottlenecks of current devices: 1) limited qubit counts (see Fig.~\ref{fig1:motivation}a), and 2) noise.

Adapting COPs to the limitations of NISQ hardware requires minimizing qubit usage. Two primary approaches have emerged: problem decomposition methods~\cite {acharya2025decomposition,zhou2023qaoa}, which partition problems into independent sub-units, and efficient qubit encoding~\cite{tan2021qubit, huber2024exponential}, which compresses classical variables into fewer quantum qubits. A significant limitation of these methods, however, is their tendency to discard correlations between variables. For complex problems such as MaxCut or portfolio optimization, these correlations shape the optimization landscape; ignoring them inevitably degrades solution quality and fails to capture the problem's intrinsic structure.


Notably, some techniques for scalable quantum computation adopt a decomposition strategy while still preserving inter-subsystem correlations in the final reconstructed quantities, such as circuit cutting~\cite{peng2020simulating,tang2021cutqc,bravyi2016trading} and entanglement forging~\cite{eddins2022doubling}. Circuit cutting introduces cuts in the circuit to split the overall computation into fragments that can run on smaller quantum processors, and then reconstructs the target expectation values via classical post-processing. Entanglement forging, by contrast, leverages a low-rank decomposition to express a larger quantum state as a linear combination of smaller-qubit states, thereby reducing quantum resources while approximately retaining entanglement-related correlations.
Recognizing that variable dependencies in COPs can be viewed through the lens of inter-subsystem correlations in quantum states~\cite{eddins2022doubling}, we extend this principle here. In this letter, we generalize the bipartite decomposition to the multipartite decomposition via truncated higher-order singular value decomposition (T-HOSVD)~\cite{vannieuwenhoven2012new}, yielding a low‑rank but efficient approximation that partitions COPs into tractable subsystems, as illustrated in Fig.\ref{fig1:motivation}b.
Each subsystem is executed on small NISQ hardware, while their mutual correlations are represented by a trainable tensor network, whose classical parameters approximate inter‑subsystem entanglement. In this way, the ensemble of subsystems reproduces the full system behavior and retains essential correlations.

We propose the Distributed Variational Quantum Algorithm (DVQA) to address the scalability limits of COPs on NISQ devices. This hybrid framework partitions the cost Hamiltonian into sub-Hamiltonians implemented on separate quantum circuits, while inter-circuit correlations are captured by a trainable tensor network. This approach defines a distributed objective function wherein global expectation values are synthesized from local measurements and tensor contractions. Through the joint optimization of quantum circuit parameters and tensor network parameters~\cite {kingma2014adam}, DVQA enables the efficient solution of large-scale problems using resource-constrained hardware.

\begin{figure}
    \centering
    \includegraphics[width=\linewidth]{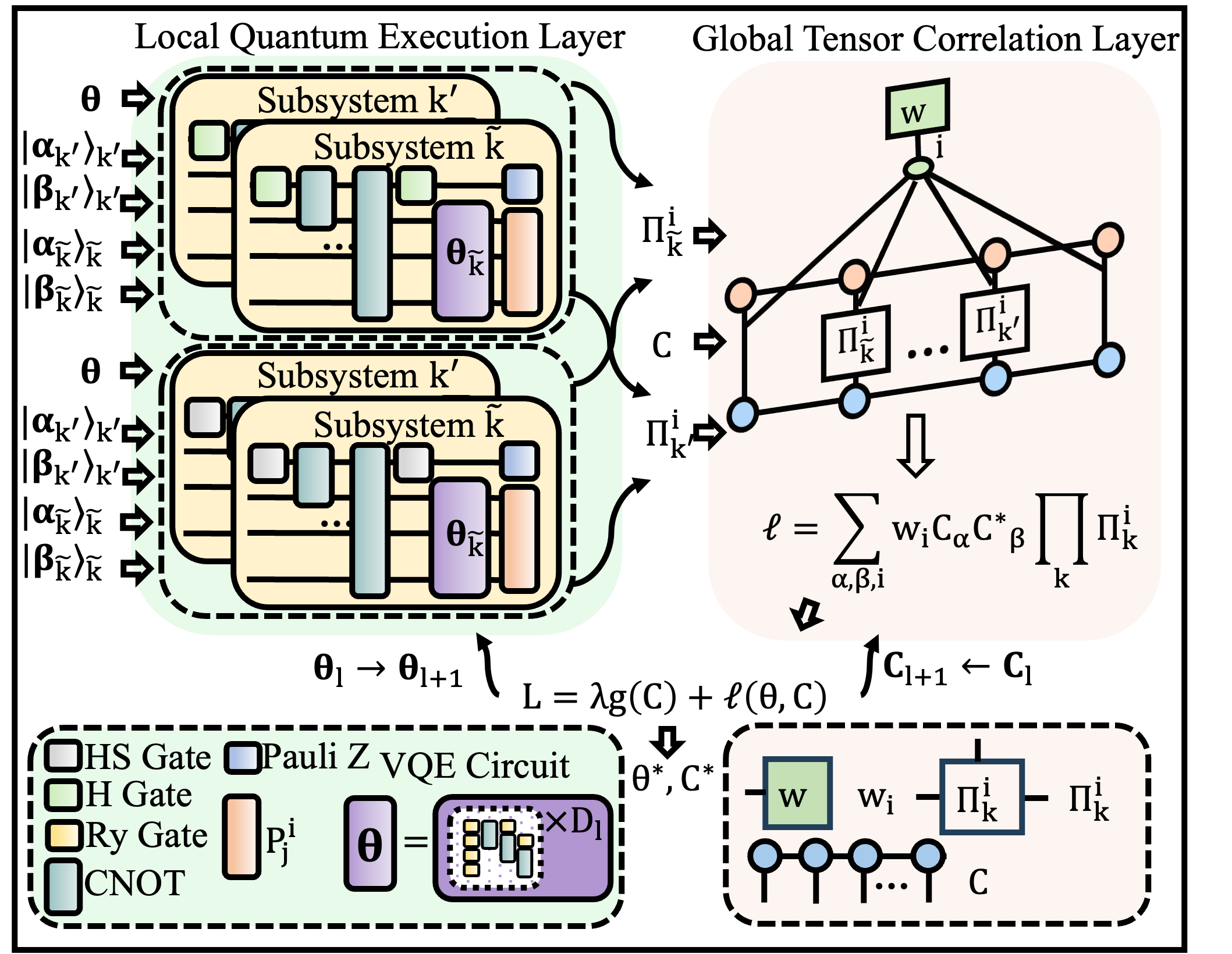}
    \caption{Schematic overview of the DVQA framework. The architecture comprises two primary modules: quantum subsystem evolution (left, green) and tensor-network-based classical optimization (right, pink). For a second-order interaction problem (e.g., QUBO p=2), each subsystem---denoted \(k'\) and \(\tilde{k}\)---has two circuits representing the real and imaginary parts of the distributed objective function. 
    Orthogonal initial states \( |\alpha_{k'}\rangle, |\beta_{k'}\rangle \) and \( |\alpha_{\tilde{k}}\rangle, |\beta_{\tilde{k}}\rangle \), together with variational parameters \( \theta=\otimes_k \theta_k \), are used to generate the circuit outputs \( \Pi_{k'}^{i} \) and \( \Pi_{\tilde{k}}^{i} \). 
    These outputs are contracted within a tensor network parameterized by trainable tensors \( C \) and weights \( w_i \), yielding the loss function $\ell(\theta, C) = \sum_{\alpha,\beta, i} w_i\, C_\alpha\, C_\beta^* \prod_k \Pi_k^i.$
    The optimization is constrained by \( C^{\dagger}C = 1 \) and performed using the ADAM optimizer through the Lagrange-multiplier formulation \( L = \lambda(C^{\dagger}C - 1) + \ell(\theta, C) \), 
    iteratively updating both $\theta$, and \(C\). 
    The final optimized parameters \( \theta_{k'}^*, \theta_{\tilde{k}}^* \), and \( C^* \) reconstruct the global quantum state $|\phi\rangle = \sum_\alpha C_\alpha^* \left( \bigotimes_{k=1}^{K} U_k(\theta_k^*)\, |\alpha_k\rangle_k \right)$,
    where the optimized subsystem states \( U_k(\theta_k^*)|\alpha_k\rangle_k \) are variationally combined through the learned tensor network.}
    \label{fig1:diag}
\end{figure}

To validate DVQA, we provide a rigorous complexity analysis proving the efficiency of the distributed objective function on NISQ systems. We then apply DVQA to two representative COPs, including MaxCut~\cite{wang2018quantum,sankar2024benchmarking} and portfolio optimization (PO)~\cite{rebentrost2024quantum,aguilera2024multi,hegade2022portfolio}, where simulations demonstrate state-of-the-art accuracy at scales up to 1,000 variables. We further substantiate the method's noise resilience through theoretical bounds on solution quality under depolarizing noise. These findings are corroborated by hardware experiments on Origin Quantum’s “Wu Kong” processor~\footnote{The Wu Kong quantum computing platform, developed by Origin Quantum, is a superconducting quantum processor accessible through a cloud interface.}, which solved a 20-dimensional PO instance with $87.7\%$ success.

Our DVQA framework provides a scalable and robust approach to NISQ-era optimization.
Our main contributions are: (i) capturing inter-subsystem entanglement via a trainable tensor network, (ii) outperforming conventional variational methods on our benchmarks, and (iii) demonstrating stability on both simulated and physical hardware.  These capabilities position DVQA as a pivotal step toward realizing large-scale combinatorial optimization on near-term quantum platforms.


\emph{\textbf{T-HOSVD Framework.--}} We first formulate the computational model and derive the distributed objective function. In a standard computational model~\cite{peng2020simulating, bravyi2016trading}, we start from a initial state $|\psi_0\rangle = |0\rangle^{\otimes N}$, where $N=n_v$ denotes the problem size. A parametrized quantum circuit $U(\theta)$, consisting of a sequence of qubit gates, then prepares the variational state $|\psi(\theta)\rangle = U(\theta)|\psi_0\rangle$. A candidate classical solution x is obtained by measuring $|\psi(\theta)\rangle$ in the computational basis. The variational objective is to minimize the expectation value of a problem‑dependent observable $H$ via optimizing $\theta$:
\begin{align}
    \ell(\theta)=\langle \psi_0|U^\dagger(\theta)HU(\theta)|\psi_0\rangle \label{model}.
\end{align}
However, as the system size $N$ increases, evaluating $\ell(\theta)$ on NISQ hardware becomes infeasible due to limited qubit connectivity, noise, and circuit depth. 

To distribute the quantum workload, we decompose the Hamiltonian into subsystem‑wise components. The Hamiltonian $H$ can be a weighted sum of tensor‑product operators,
\begin{align}
    H = \sum_{i=1}^M w_i h_i, \quad h_i = \bigotimes_{k=1}^K P_k^i, \label{Eq: ham_decom}
\end{align}
where $P_k^i \in \{I, X, Y, Z\}^{\otimes n_k}$ represents local Pauli operators acting on the $n_k$ qubits of subsystem $k$, and $\sum_{k=1}^K n_k = N$. This representation isolates local interactions and provides a basis for distributed quantum evaluation on multiple devices. 

With the system partitioned into $K$ subsystems, any state on the full Hilbert space can be written as:
\begin{align}
    |\psi\rangle = \sum_{i_1, i_2, \dots, i_K} X_{i_1 i_2 \dots i_K} |i_1\rangle_1 \otimes |i_2\rangle_2 \otimes \cdots \otimes |i_K\rangle_K,
\end{align}
where $|i_k\rangle_k=\{0,...,2^{n_k}-1\}$ represents a basis state in the $k$-th subsystem, and $X_{i_1 i_2 \dots i_K}$ are the complex amplitudes. The tensor $X$ represents the correlations among subsystems. However, directly handling $X$ requires operations that scale exponentially with $K$, limiting scalability. To obtain a tractable yet expressive approximation, we apply the HOSVD to $X$ (details in Supplementary.~S1), obtaining physical state:
\begin{align}
    |\psi\rangle = \sum_{\alpha} \Tilde{C}_\alpha \left( \bigotimes_k U_k (\theta_k) |\alpha_k\rangle_k \right), \quad \sum_{\alpha} \Tilde{C}_\alpha^2 = 1, \label{Eq: state decom}
\end{align}
where $\alpha = \{\alpha_1, \dots, \alpha_{K}\}$, with $\{\alpha_k\}_k \in \{0,1, \dots, R_k-1\}$, representing computational basis states in the $k$-th subsystem, $R_k$ is the rank determined by T-HOSVD decomposition, and $\Tilde{C}$ is amplitude tensor. Simulating this quantum state generally requires inter-subsystem entanglement. 
However, under NISQ resource constraints, generating such entanglement can be prohibitive. 
Instead, we use a classical parameter tensor $C$ to approximate the resulting inter-subsystem correlations and thereby capture the dominant quantum effects. In this representation, the virtual state is: $|\phi\rangle = \sum_{\alpha} C_\alpha \left( \bigotimes_k U_k (\theta_k) |\alpha_k\rangle_k \right)$. This representation retains global expressivity through the classical tensor $C$ while enabling distributed evaluation on isolated NISQ devices, shown in Fig.~\ref{fig1:motivation}b).

Substituting Eqs.~(\ref{Eq: ham_decom}) and (\ref{Eq: state decom}) into Eq.~(\ref{model}) yields the distributed objective function:
\begin{align}
  \ell(C, \theta) &= \sum_i w_i \sum_{\alpha, \beta} C_\alpha C_\beta^* \prod_k \text{Tr}\left(P_k^i U_k(\theta_k) |\alpha_k\rangle_k \langle \beta_k|_k U_k^\dagger(\theta_k)\right),  \\
  &=\sum_i w_i \sum_{\alpha, \beta} C_\alpha C_\beta^* \prod_k \Pi^i_k
  \label{eq: L}
\end{align}
where $\Pi^i_k$ denotes the local‑expectation term measured on subsystem $k$. Each term in the product can be independently measured on a small quantum circuit associated with subsystem $k$, while classical tensor‑network contractions reconstruct the global expectation. This formulation defines the theoretical core of DVQA (details in Supplementary.~S2): a hybrid objective function that distributes quantum workloads and captures global correlations through a classical tensor network. 


\emph{\textbf{DVQA Implementation.--}}
To realize the DVQA in near-term hardware, we establish a hybrid quantum–classical workflow that proceeds through five stages: (1) quantum‑subsystem circuit execution and measurement, (2) classical tensor‑network aggregation, (3) computational‑complexity evaluation, (4) noise robustness analysis, and (5) joint optimization of quantum and classical parameters. Figure~\ref{fig1:diag} gives a schematic overview of the entire pipeline.


\emph{Quantum implementation.}
The green block in Fig.~\ref{fig1:diag} represents the independent quantum subsystems described by a parameterized circuit $U_k(\theta_k)$ (shown as the purple block inside each subsystem), composed of layers of single‑qubit rotations and CNOT entanglers in a hardware‑efficient ansatz (HEA)~\cite{kandala2017hardware}. For each subsystem, we evaluate the local complex quantities $\Pi_k^i = \langle \alpha_k | P_k^i | \beta_k \rangle$. To access real and imaginary components, we employ ancilla‑assisted circuits inspired by the Hadamard‑test interference scheme~\cite{bravyi2016trading}. An ancilla qubit is prepared in a superposition and entangled with the subsystem through a controlled‑$U$ operation, where the internal unitary $U$ maps the reference state $|\alpha_k\rangle \!\to\! |\beta_k\rangle$. Measuring the ancilla in the $Z$ basis yields the real part of the expectation value, $\operatorname{Re}{\langle \alpha_k | P_k^i | \beta_k \rangle}$, whereas inserting an additional phase‑shift gate ($S$) after the Hadamard gate rotates the ancilla basis to obtain the imaginary part. The subsystem circuit further incorporates the parameterized ansatz $U_k(\theta_k)$ before measurement of the observable $P_k^i$. Complete circuit diagrams and the proof of the circuit design are provided in Supplementary~S3. The resulting expectation values ${\Pi_k^i}$ are elements of a tensor that represent the outcomes of quantum measurements. These values are then communicated to the classical aggregation stage described below.


\emph{Classical tensor contraction.}
In the classical stage, depicted in Fig.~\ref{fig1:diag} pink block, all local measurement results are combined through a tensor‑network contraction parameterized by a trainable tensor $C$. Each tensor element $C_\alpha$ corresponds to the coefficient of the composite basis state $|\alpha\rangle$, encoding correlations among subsystems. The distributed objective function is therefore $\ell(C, \theta) = \sum_i w_i \sum_{\alpha, \beta} C_\alpha C_\beta^* \prod_k \Pi^i_k$, where $w_i$ are coefficients in the target Hamiltonian. This contraction integrates subsystem information purely at the classical level, avoiding
inter‑subsystem entangling operations while preserving global correlations across subsystems. The tensor network construction and contraction order for efficient implementation are provided in Supplementary~S4.

\emph{Computational complexity.}
Next, we specify the parameters relevant to resource scaling. Here $M$ denotes the number of Hamiltonian terms, $K$ the number of subsystems, $m$ the bond dimension of each tensor composing $C$, $R_{\max}$ the highest rank among subsystems, and $p$ the locality order of subsystem interactions. 
The quantum-sampling complexity required to estimate $\ell(C,\theta)$ within additive error
$\varepsilon$ is $O\!\left(\tfrac{\lVert w\rVert_1^2 \lVert C\rVert_1^4}{\varepsilon^2}\right)$, where
$\|w\|_1=\sum_{\alpha}|w_{\alpha}|$ and $\|c\|_1=\sum_{\beta}|c_{\beta}|$ are the $\ell_1$ norms
of the Hamiltonian coefficient vector $w$ and the tensor-network vector $C$, respectively.
Because the tensor network parameters are normalized ($\sum_\alpha |C_\alpha|^2=1$), the norm obeys $\lVert C\rVert_1 \le \sqrt{\mathcal{D}}$ with $\mathcal{D}$ the number of tensor elements. When the bound dimension $m$ and $R_{\max}$ scale at most polynomially with the system size $N$, $\mathcal{D}$ is also polynomial, ensuring that the sampling cost remains polynomially bounded and thus tractable. 
The number of quantum‑circuit variants scales as $O(MR_{\max}^{4p})$, and the corresponding classical tensor‑contraction workload is $O(KR_{\max}Mm^3 + pR_{\max}^2Mm^2)$. In many relevant problems, such as quadratic unconstrained binary optimization (QUBO, $p\!=\!2$) and 3‑SAT ($p\!=\!3$), the locality $p$ is small and both $m$ and $R_{\max}$ are observed to scale at most polynomially with $N$. Under these realistic conditions, the quantum‑sampling, circuit‑execution, and classical‑contraction stages all scale polynomially, confirming that DVQA can be efficiently implemented on NISQ‑era devices. Full derivations are provided in Supplementary.~S5.


\emph{Noise robustness.}
To evaluate the experimental stability of DVQA, we model each qubit as undergoing an independent and identically distributed (i.i.d.) depolarizing‑noise channel~\cite{urbanek2021mitigating}, denoted by $\mathcal{E}(\cdot)$, with single‑ and two‑qubit error probabilities $p_1$ and $p_2$, respectively. For a subsystem circuit whose HEA has depth $D_l$ and acts on $d$ qubits, we define the noise-induced expectation-value error as
\begin{equation}
\Delta H \equiv \mathrm{Tr}[H\,\mathcal{E}(\rho)] - \mathrm{Tr}[H\rho].
\end{equation}
Under the i.i.d.\ noise assumption, the magnitude of this error can be approximately bounded as
\begin{align}
|\Delta H|_{\text{i.i.d}}
\leq (1-(f_{\mathrm{sub}})^p) \, |\mathrm{Tr}[H\rho]|.
\label{eq:noise_decay}
\end{align}
where $f_{sub}=(1-p_1)^{(4 + D_l d)}
(1-p_2)^{(\log R_{\max} + D_l(d-1))}$, and $p$ denotes the interaction locality of the Hamiltonian $H$. As shown in Eq.~(\ref{eq:noise_decay}), the impact of noise is governed by the interaction order $p$ through the factor $(f_{\mathrm{sub}})^p$, where $f_{\mathrm{sub}}$ depends on the circuit depth $D_l$, the subsystem size $d$, the single- and two-qubit error probabilities $p_1$ and $p_2$, and $\log R_{\max}$, rather than the total qubit number $N$.

While Eq.~(\ref{eq:noise_decay}) assumes i.i.d. depolarizing noise, real NISQ devices often exhibit correlated or non‑i.i.d. noise due to crosstalk, drift, or gate‑dependent errors. 
By treating these correlations as a small perturbation ($\epsilon_c \!\ll\! 1$), the first‑order correction to the expectation value can be bounded as
\begin{align}
|\Delta H|_{\text{non‑i.i.d.}}
\leq
(f_{\mathrm{sub}})^{p-1} p C_B D_l \epsilon_c,
\end{align}
where $f_{\mathrm{sub}}$ is the i.i.d. attenuation factor, $p$ denotes the interaction locality, $D_l$ the depth of the quantum circuit for each subsystem, and $C_B$ a constant. 
Overall, the total noise deviation can be bounded by
\begin{align}
|\Delta H|_{\text{total}}
\leq
(1 - (f_{\mathrm{sub}})^p)\,|\mathrm{Tr}[H\rho]|
+
(f_{\mathrm{sub}})^{p-1} p C_B D_l \epsilon_c,
\end{align}
indicating that, for physically relevant Hamiltonians featuring low‑order interactions ($p\!\le\!3$), DVQA remains robust under realistic correlated‑noise conditions.
Detailed derivations and the full noise‑propagation analysis can be found in Supplementary.~S6.


\emph{Joint optimization and final state.}
The optimization involves two distinct parameter types: quantum circuit parameter $\theta$ and a trainable tensor $C$. The gradient of quantum circuit parameter $\theta_g$ of the g-th system is (Details in Supplementary.~S7.1):
\begin{align}
    \nabla_{\theta_g}L = \sum_{i, \alpha, \beta} w_i C_\alpha C_\beta^* & \Bigg( \prod_{j \neq g} \Pi(\alpha_j,\beta_j,\theta_j)^i_j \frac{\delta \Pi(\alpha_g,\beta_g,\theta_g)^i_g}{\delta \mathbf{\theta}} \Bigg).
\end{align}
To fulfill the normalization constraint of a trainable tensor $C$, we employ Lagrange multipliers with Lagrangian $L = \ell(C, \mathbf{\theta}) + \lambda (C^T C - 1)$. The gradient of $C$ is constrained to the tangent space of the hypersphere, yielding the gradient of $C$ (details in Supplementary.~S7.2):
\begin{align}
    \nabla_C L = \nabla_C \ell(C, \mathbf{\theta}) - 2\lambda C,
\end{align}
where $\lambda = \frac{C^T \cdot \nabla_C \ell(C, \mathbf{\theta})}{2}$ ensures the constraint is maintained throughout optimization. The distributed objective function is optimized using the ADAM optimizer, with the gradients of $\theta$ and $C$.

After optimization, the final result is obtained from the reconstructed final state $|\phi\rangle$ by combining the subsystem states:  
\begin{align}
    |\phi\rangle = \sum_\alpha C^*_\alpha \bigotimes_{k=1}^K U_k(\theta^*_k) |\alpha_k\rangle_k,
\end{align}
where $\theta^*$ and $C^*$ are the optimized parameters. Using these parameters, the final solution to the problem is determined, and the corresponding loss value is evaluated.

\emph{\textbf{Applications to COP Problems.--}} 
In this section, we apply the DVQA to representative combinatorial optimization problems, specifically the MaxCut and Portfolio Optimization tasks, to demonstrate its accuracy, scalability, and robustness to noise. 

\emph{DVQA framework for COPs.}
DVQA reformulates combinatorial optimization problems into distributed quantum objectives that are optimized jointly across classical and quantum subsystems. Each subsystem represents a local variational quantum state, and global correlations are captured through a tensor-network interconnection with a tunable rank, set by the T-HOSVD approximation. This distributed formulation, inspired by the entanglement‑forging framework, enables decomposition of a large optimization landscape into smaller, hardware‑compatible fragments.

In most combinatorial optimization problems, such as the QUBO model, the optimal configuration corresponds to a single computational‑basis state or a superposition of a few low‑rank basis configurations. The rank‑tunable architecture of DVQA naturally matches this structure. The increasing rank $R_{max}$ extends the expressivity of DVQA and smooths the optimization landscape, leading to faster convergence (see Supplementary.~S8).

According to the theoretical noise bound (Eq.~(\ref{eq:noise_decay})), the accumulation of noise in DVQA depends mainly on the local circuit depth $D_\ell$, subsystem size $d$, and the interaction order $p$ of the target Hamiltonian. Most practical COP Hamiltonians, such as MaxCut and Portfolio Optimization tasks, are low‑order (typically 2‑local), leading to relatively moderate noise amplification. As a result, DVQA maintains high accuracy and robustness when applied to such systems, even on NISQ‑era hardware.


Therefore, DVQA is particularly well suited for optimization problems whose ground states exhibit low entanglement dimensionality—and can therefore be efficiently represented within low‑rank subspaces~\cite{zhang2024analyzing}—as well as those governed by low‑order Hamiltonians, precisely the structural characteristics of many combinatorial optimization problems.

\emph{Problem mappings.}
The MaxCut (details in Supplementary.~S9.1) and Portfolio Optimization (details in Supplementary.~S9.2) problems are reformulated into equivalent quantum representations by mapping them to Ising Hamiltonians. For MaxCut, the objective is to maximize the total weight of edges between two subsets. This can be expressed with the Hamiltonian:
\begin{equation}
H = -\sum_{(i,j)\in E} \frac{w_{ij}(1 - Z_i Z_j)}{2},
\end{equation}
where $E$ denotes the set of edges $(i,j)$ in the problem graph, and each $w_{ij}$ represents the corresponding edge weight. 
The operators $Z_i$ and $Z_j$ are Pauli‑$Z$ operators acting on the $i$‑th and $j$‑th qubits, whose eigenvalues $\pm1$ encode the binary variables of the classical MaxCut formulation.
Each interaction term $Z_i Z_j$ corresponds to an edge $(i,j)$, such that minimizing the expectation value $\langle H \rangle$ reproduces the objective of maximizing the total edge weight between partitions.
For the PO problem, which aims to balance maximizing return and minimizing risk, the corresponding Hamiltonian is:
\begin{equation}
H = \sum_{i} h_i Z_i + \sum_{i \neq j} J_{ij} Z_i Z_j + C,
\end{equation}
where the 
$h_i = \frac{1}{2} \left[ (1-\lambda) r_i - \lambda \sum_{j} V_{ij} +\frac{\lambda}{2}V_{ii}\right]$, 
$J_{ij} = \frac{\lambda}{4} V_{ij}$,
and $C = \frac{\lambda}{4} \sum_{ij} V_{ij} - \frac{1-\lambda}{2} \sum_i r_i$. The $\lambda$ is the risk tolerance parameter, $V$ is the risk matrix, and $r$ is the return vector. 

\emph{Evaluation Protocol.}
To compare with other approximate algorithms, we use the metric called the approximation ratio. Since in our setting both $\text{OPT}(I)$ and $A(I)$ are negative, for a minimization problem, we define the approximation ratio as
\begin{equation}
    \rho_{\text{min}} = \frac{\text{OPT}(I)}{A(I)}.
\end{equation}
where $A(I)$ is the value obtained by a method, and $\text{OPT}(I)$ is the optimal value.

\begin{figure}[htpb]
\centering
\includegraphics[width=0.5\textwidth]{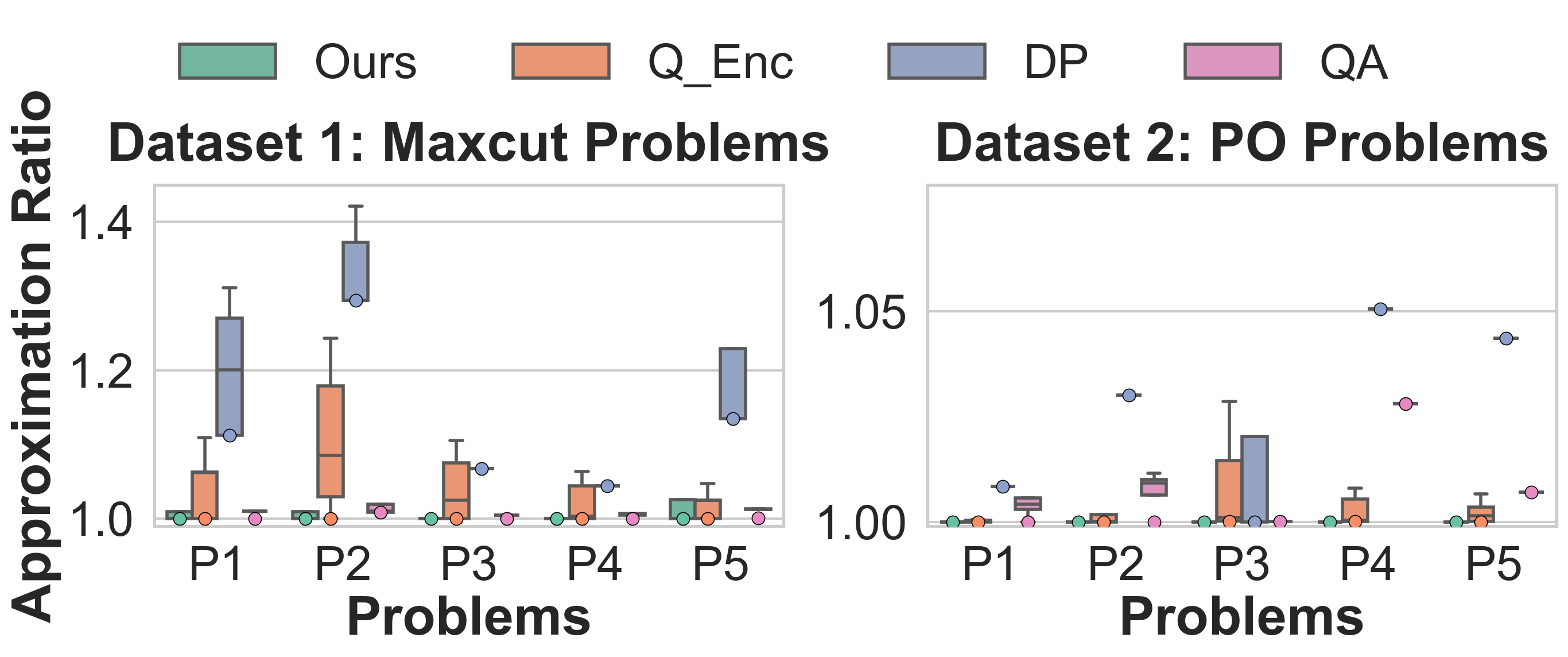}
\caption{Performance comparison of different optimization methods on MaxCut (left, Dataset~1) and portfolio optimization (right, Dataset~2). The vertical axis shows the approximation ratio and the horizontal axis lists problem instances (P1--P5); colors denote different methods (Ours, Q\_Enc, DP, QA). For each problem--method pair, we perform 20 independent runs. We use a box-and-whisker plot with outliers suppressed: the box spans the interquartile range (IQR) from the first quartile $Q_1$ (25th percentile) to the third quartile $Q_3$ (75th percentile), the center line indicates the median $Q_2$ (50th percentile), and the whiskers extend to the most extreme observations within $[Q_1-1.5\,\mathrm{IQR},\,Q_3+1.5\,\mathrm{IQR}]$, where $\mathrm{IQR}=Q_3-Q_1$. In addition, colored circular markers denote the run with the best objective value among the 20 runs for each problem--method pair.}
\label{fig:sota}
\end{figure}

\emph{Performance and Scalability Results.}
DVQA is compared with the Quantum Annealing (QA)~\cite{morrell2024quantumannealing}, Qubit Encoding (Q\_Enc)~\cite{tan2021qubit}, and Decomposition Pipeline (DP)~\cite{acharya2025decomposition}, as described in Supplementary.~S10, performing on two types of optimization problems: the MaxCut problem and the PO problem (S\&P dataset and CoinGecko dataset). 
Each method is executed independently 20 times on each problem instance under consistent resource constraints. We use an optimization budget of 200 iterations and a circuit depth of 6 layers for Ours, Q\_Enc, and DP, and approximately 200 optimization steps for QA.
The approximation ratios $\rho_{\text{min}}$ obtained by different methods are summarized in Fig.~\ref{fig:sota}, 
where the x-axis represents the five problem instances and the y-axis shows the approximation ratio.
For both MaxCut and portfolio optimization, DVQA performs comparably to or
better than QA, Q\_Enc, and DP across all tested instances, consistently
achieving approximation ratios close to $1$. These results indicate that DVQA is overall superior to state-of-the-art baseline methods for solving combinatorial optimization problems. To further understand how the subsystem dimension $d$ and the bound dimension $m$ of $C$ affect the optimization performance of DVQA, we provide detailed ablation studies in Supplementary.~S11.

\begin{figure}[htpb]
\centering
\includegraphics[width=0.5\textwidth]{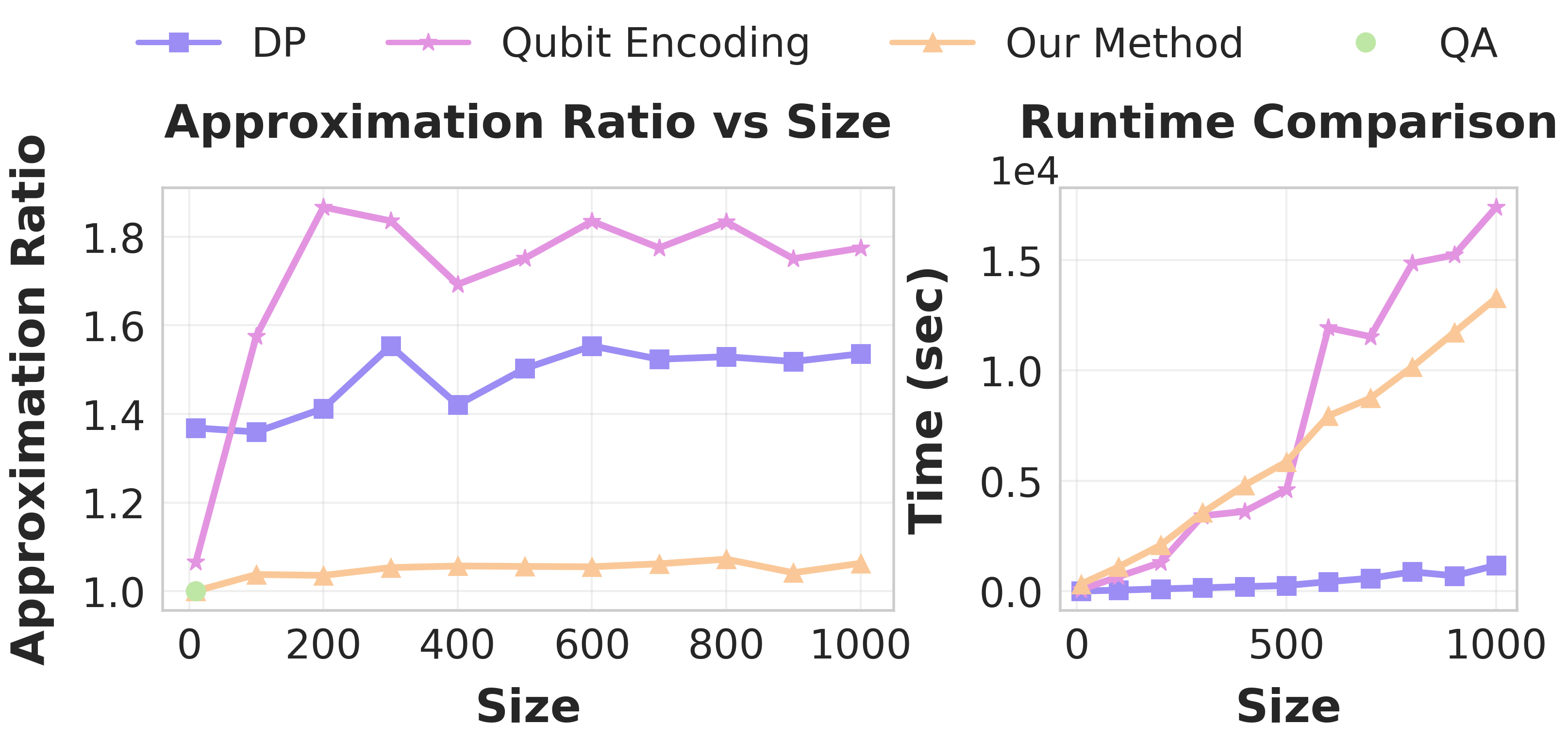}
\caption{Scalability (left) and runtime (right) comparison of different methods. The left panel shows the approximation ratio versus problem size (Size), and the right panel shows the end-to-end runtime (Time, in seconds) versus Size.}
\label{fig:sca}
\end{figure}

\begin{figure*}[htpb]
\centering
\includegraphics[width=\textwidth]{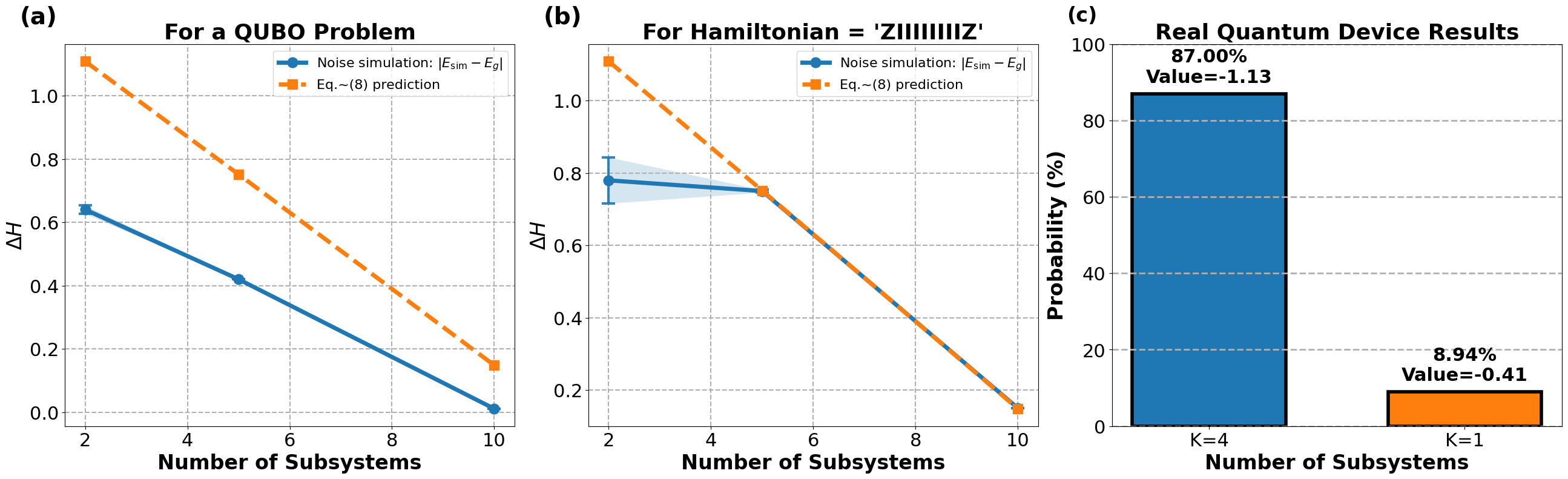}
\caption{
Verification of the DVQA noise model and real-device performance.
(a) For a representative QUBO instance ($p=2$), comparison between the noise-simulation results and the theoretical prediction of Eq.~(8) as a function of the number of subsystems $K$.
The blue markers report $\Delta H = \left|E_\text{sim} - E_g\right|$ averaged over 10 independent runs; the error bars and the shaded band both indicate standard deviation (STD) across the 10 runs.
The orange squares show the Eq.~(8) prediction.
(b) Same as (a) but for the two-body Hamiltonian $H=\mathrm{ZIIIIIIIIZ}$; results are averaged over 10 independent runs, with error bars and shaded bands denoting STD.
(c) Real quantum device results on the \textit{Origin Quantum Wu--Kong} superconducting processor for the SP20 problem, comparing DVQA ($K=4$) with the original VQE setting ($K=1$).
Bars indicate the measured success probability, and the annotated values report the corresponding energy.
}
\label{fig:exp}
\end{figure*}

To evaluate the scalability and runtime of DVQA, a simulation analysis, shown in Fig.~\ref{fig:sca}, is conducted on MaxCut problems with dimensions ranging from 10 to 1000, assuming $R_{\text{max}} = 1$ and a subsystem size of 6 qubits. Four optimization methods are compared based on their approximation ratio and runtime. The approximation ratio results, shown in the left plot, indicate that Our Method achieves the best performance, converging close to 1.05, while Q\_Enc and DP converge to approximately 1.8 and 1.5, respectively. QA is represented as a single point due to memory limitations in the simulation, which restricts its scalability. The right plot presents the runtime comparison, where Our Method exhibits polynomial growth, effectively balancing scalability and efficiency.

\emph{Noise Analysis and Real‑Device Demonstration.}
Eq.~(\ref{eq:noise_decay}) shows that the noise influence in DVQA depends mainly on the number of qubits in each subsystem ($d$), the circuit depth per subsystem ($D_\ell$), the qubit noise ($p_1$, $p_2$), rank $R_{max}$ and the interaction order ($p$). When the total number of qubits $N$ is fixed, increasing $K$ reduces the subsystem size $d$, thereby decreasing the overall noise effect, with all other parameters fixed.

Fig.~\ref{fig:exp} (a) validates the above noise-scaling behavior on a representative
$10$-variable QUBO instance with $p=2$.
We quantify the deviation from the ground-state energy $E_g$ by
$\Delta H = |E_{\mathrm{sim}} - E_g|$.
Under the single-qubit depolarizing rate $p_1=0.01$ and the two-qubit rate
$p_2=0.2$, for each number of subsystems $K$, we perform $10$ independent runs.
Markers show the mean value, and the shaded region and error bars indicate the standard deviation.
As the number of quantum subsystems increases (i.e., the subsystem size decreases), $\Delta H$ decreases, indicating a positive correlation between the error magnitude and the subsystem size in the low-$p$ regime.
This suggests that, for DVQA applied to COPs in our low-$p$ setting, the error
is dominated by within-block noise accumulation, while cross-block propagation
is comparatively limited, consistent with an effectively localized noise effect
with respect to the subsystem decomposition.
Moreover, Eq.~\ref{eq:noise_decay} captures the observed scaling of $\Delta H$ with subsystem size and consistently upper-bounds $\Delta H$ in our simulations under this noise setting.
To further probe the model in a simpler and more controlled setting, we
consider a two-body Hamiltonian $H=\mathrm{ZIIIIIIIIZ}$, as shown in
Fig.~\ref{fig:exp}(b).
Using the same noise parameters and $10$ independent runs, the prediction of
Eq.~\ref{eq:noise_decay} agrees well with the noise simulations, particularly at larger numbers of subsystems; the small discrepancy at smaller numbers of subsystems is mainly due to fluctuations from the variational optimization and the finite number of repeats, as reflected by the shaded band and error bars.
More detailed numerical simulations of the depolarizing-noise effect for different subsystem numbers and two-qubit noise rates $p_2$ are provided in Supplementary.~S12.


Finally, we test DVQA on the \textit{Origin Quantum Wu–Kong} superconducting quantum computer to examine its real hardware performance. Five high-fidelity qubits are sequentially reused across four iterations to represent the four subsystems of an SP20 problem. Each quantum circuit evaluation is executed with $N_{\text{shot}} = 1000$ measurement shots. Each subsystem reaches its correct solution with success probabilities of 94.33\%, 99.2\%, 95.01\%, and 98.62\%, respectively, and the combined global solution achieves an overall success rate of about 87.7\% (Fig.~\ref{fig:exp}(c)). Compared to the original VQE case ($K=1$), DVQA produces a lower objective energy value and a significantly higher success probability, demonstrating its superior robustness and performance under realistic noise conditions.

\emph{\textbf{Summary.--}}
Current quantum optimization strategies often compromise solution quality to fit large problems onto limited NISQ hardware. In this Letter, we propose the DVQA, a hybrid quantum–classical framework that scales COP optimization on NISQ devices by distributing the workload across many small subsystems while preserving inter-variable correlations via a tunable low-rank T-HOSVD reconstruction, thereby naturally localizing noise.
This architecture enables the processing of 1,000-variable instances, achieving state-of-the-art accuracy where traditional decomposition methods fail.
Importantly, for Hamiltonians with low interaction order $p$, DVQA exhibits favorable noise scaling on NISQ hardware, with error accumulation localized at the subsystem level.
Unlike existing problem-decomposition and qubit-encoding methods that treat subsystems as nearly independent, DVQA captures cross-subsystem correlations using a trainable tensor network, enabling scalable and noise-resilient optimization on near-term quantum processors.

\emph{\textbf{Acknowledgments.--}} This work acknowledges support from the Hong Kong Research Grants Council (RGC) under Grant No. R6010-23, as well as support from HSBC.


\bibliography{apssamp}

\end{document}